\documentclass[pre,preprint,showpacs,aps]{revtex4-1}

\usepackage{graphicx}
\usepackage{amsmath}

\begin{document}

\title{Lyapunov Exponent and Criticality in the Hamiltonian Mean Field Model}

\author{L.~H.~Miranda Filho}
\affiliation{Departamento de F\'\i sica, Universidade Federal Rural de Pernambuco,
Rua Manoel de Medeiros, s/n - Dois Irm\~aos, 52171-900 - Recife, Brazil}
\author{M.~A.~Amato}
\affiliation{Instituto de F\'{i}sica and International Center for Condensed Matter Physics,
Universidade de Bras\'\i lia, CP 04455, 70919-970 - Bras\'\i lia, Brazil}
\author{T.~M.~Rocha Filho}
\email{marciano@fis.unb.br}
\affiliation{Instituto de F\'{i}sica and International Center for Condensed Matter Physics,
Universidade de Bras\'\i lia, CP 04455, 70919-970 - Bras\'\i lia, Brazil}

%\date{} 

\begin{abstract}
We investigate the dependence of the largest Lyapunov exponent of a $N$-particle self-gravitating ring model at equilibrium
with respect to the number of particles and its dependence on energy. This model has a continuous phase-transition from a ferromagnetic
to homogeneous phase, and we numerically confirm with large scale simulations the
existence of a critical exponent associated to the largest Lyapunov exponent,
although at variance with the theoretical estimate. The existence of strong chaos in the magnetized state evidenced by a positive Lyapunov exponent
is explained by the coupling of individual particle oscillations to the diffusive motion of the
center of mass of the system and also results on a change of the scaling of the largest Lyapunov exponent with the number of particles.
We also discuss thoroughly for the model the validity and limits of the approximations made by a geometrical model
for their analytic estimate.
\end{abstract}

\maketitle

\section{Introduction}

Many body systems with long range interactions are known to have several properties that set them apart
from more ``usual'' systems with short range interactions, such as ensemble inequivalence,
negative heat capacity (with no second law violation), anomalous diffusion and
non-Gaussian (quasi-) stationary states~\cite{booklri}. An interparticle interaction potential
is said to be long ranged if it decays at large distances as $r^{-\alpha}$ with $\alpha \leq d $,
$d$ the spatial dimension, with a consequence that 
the total potential energy increasing superlinearly with volume~\cite{campa,dauxois}.
Some important physical system with long range interactions are
non-neutral plasmas~\cite{rizzato}, self-gravitating systems~\cite{tanu}, vortices in two-dimensional turbulent hydrodynamics~\cite{venaille}
and free electron laser~\cite{anton1}. Simplified models were also largely considered in the literature and allowed a better understanding
of the statistical mechanics of equilibrium and non-equilibrium of systems with long range interactions, such as
one and two-dimensional self-gravitating systems~\cite{mila,teles}, the Hamiltonian Mean Field (HMF) and self-gravitating ring models~\cite{ruffoHMF,sota}.

Much progress in the understanding of the relaxation properties in many-particle systems with long range forces
came from numerical simulations of model systems~\cite{steiner,rl1,rl2,rl3,rl4,rl5,rl6,rl7,rl8}.
Although the scaling of the relaxation time to equilibrium with $N$ depends
on the type of system and spatial dimension~\cite{scaling,chris}, as a common feature it diverges with $N$, and as a consequences it never attains
thermodynamic equilibrium for $N\rightarrow\infty$. In many cases this relaxation time is sufficiently large that even for finite $N$ it
can be considered infinite for practical purposes. If the equilibrium is reached, then its properties can be studied using the usual
techniques of equilibrium statistical mechanics~\cite{booklri,campa,dauxois,levin}.

Simplified models have been important in the study of the intricate interplay between chaotic dynamics, ergodic properties and statistical mechanics
of systems with long range interactions,
while Lyapunov exponents has proven to be a useful tool in the study of chaos in dynamical systems~\cite{ott}
and particularly also for long range systems~\cite{vallejos1,firpo1,anteneodo1}.
The precise determination of Lyapunov exponents is an intricate task and usually requires a great numerical effort with very long integration times,
that can become prohibitive for a system with a very large number of particles. A prescription for their analytic estimation is therefore
of great relevance. Casetti, Pettini and collaborators developed an analytic method
to obtain the scaling behavior of the Largest Lyapunov Exponent (LLE)~\cite{pettini1,pettini2,pettini3}, and applied to
the HMF model by Firpo in Ref.~\cite{firpo2}.

The HMF model has been widely studied in the literature as a prototype for observations of some dynamical features of long range
interacting systems~\cite{booklri}. It consists of $N$ classical particles moving on a unit circle and globally coupled with
Hamiltonian~\cite{ruffoHMF}:
\begin{equation}
H=\sum_{i=1}^{N} \frac{p_i^2}{2m}+\frac{1}{2N}\sum_{i,j=1}^{N}\left[1-\cos{(\theta_i - \theta_j)}\right],
\label{hmf}
\end{equation}
where $\theta_i$ and $p_i$ are the position angle on the circle and $p_i$ its conjugate momentum. The $1/N$ factor in the potential energy
is the Kac factor introduced such that the total energy is extensive, and can be obtained from a change in the time unit.
We can define by analogy the total magnetization and its components by:
\begin{equation}
{\bf M}=\left(M_{x},M_{y}\right)= \frac{1}{N} \sum_{i=1}^{N}(\cos{\theta_{i}},\sin{\theta_{i}}).
\label{mag}
\end{equation}
The equations of motion are then
\begin{eqnarray}
\dot\theta_i & = & p_i,
\nonumber\\
\dot p_i & = & -\sin\theta_i\:M_x+\cos\theta_i\:M_y.
\label{HMFeqmot}
\end{eqnarray}
As a thermodynamic system this system is exactly solvable, i.~e.\ its equilibrium partition function is obtained in closed form,
and its equilibrium distribution function is given by~\cite{ruffoHMF}:
\begin{equation}
f_{\rm eq}(p,\theta)=\frac{\sqrt{\beta}}{(2\pi)^{3/2}\:{\rm I}_0(\beta)}e^{-\beta\left(p^2/2-M\cos(\theta)\right)},
\label{eqdisthmf}
\end{equation}
where ${\rm I}_k$ is the modified Bessel function of the first kind with index $k$, and the origin for angles is chosen such that
$M_y=0$ and $M_x=M$. For a given inverse temperature $\beta$, the magnetization is obtained
from the equation:
\begin{equation}
M=\frac{{\rm I}_1(\beta M)}{{\rm I}_0(\beta M)},
\label{eqvalM}
\end{equation}
with $M\equiv||{\bf M}||$.
The dependence of $M$ on temperature is thus obtained by solving Eq.~(\ref{eqvalM}). A second order phase transition occurs at the
critical energy per particle $e_c=E_c/N=3/4$ and $T=1/\beta=0.5$ from a lower energy
ferromagnetic phase to a higher energy phase with zero magnetization.
Canonical or microcanonical ensembles are fully equivalent for the HMF model.
Out of equilibrium phase transitions for this model were studied in some detail in~\cite{amato3}.

In the present work we investigate the applicability of the geometrical approach of Refs~\cite{pettini1,pettini2} by
directly testing its underlying assumptions. We also consider the scaling with $N$ of the LLE for the HMF model at different energy ranges,
and compare our numerical results to theoretical results and other similar numerical investigations,
for larger values of $N$ than in previous studies.
Particularly we confirm the existence of a new critical exponent
corresponding for the LLE theoretically predicted in~\cite{firpo2} although with a small deviation from the predicted value of the exponent.

This paper is structured as follows: in section~\ref{sec2} we briefly recall the theory of Lyapunov Exponents and
the numerical determination of the LLE.
In section~\ref{sec3} we present and discuss our results for the HMF model and we close in section~\ref{sec4} with some concluding remarks.

\section{Lyapunov Exponents}
\label{sec2}

A Lyapunov Exponent (LE) quantify how the dynamics of the system is sensible to small differences in the initial conditions.
With this aim, let us define the vector formed by coordinates in a $n$-dimensional phase space:
\begin{equation}
\bf{x}\equiv(x_1,x_2,....x_n).
\label{phase coord}
\end{equation}
which we suppose satisfy a set of $n$ autonomous first-order differential equations:
\begin{equation}
\frac{d\bf{x}(t)}{dt}={\mathbf F}(\bf{x}(t)).
\label{flow}
\end{equation}
Equation~(\ref{flow}) generates a flows in the phase space, and $F(\bf{x}(t))$ is the velocity field of the flow.
In order to measure contraction or stretching in the neighborhood of $\bf{x}(t)$, we consider two different
solutions of Eq.~(\ref{flow}) $\bf{x}^{(1)}(t)$ and $\bf{x}^{(2)}(t)$ and the difference vector
$\mathbf{w}\equiv{\bf{x}}^{(2)}(t)-{\bf{x}}^{(1)}(t)$:
\begin{equation}
\mathbf{w}=(\delta x_1, \delta x_2, \dots, \delta x_n).
\label{dev vec}
\end{equation}
The evolution equation for $\mathbf{w}$ is then:
\begin{equation}
\frac{d{\mathbf{w}}}{dt} = \mathbf{J}(\mathbf{x}(t))\mathbf{w},
\label{flow linear}
\end{equation}
with $\bf{J}$ the $N\times N$ Jacobian matrix of the flow. Assuming that the elements of $\bf{J}$ are continuous bounded functions of $t$
for $t\rightarrow\infty$, then the solutions of  ($\ref{flow linear}$) grow no faster than $\exp(\lambda t)$, for some constant $\lambda$.

The Lyapunov Exponent for a given initial condition $\mathbf{w}(0)$ is defined by
\begin{equation}
\lambda\equiv\displaystyle \lim_{t \to \infty} \frac{1}{t} \ln\left(\frac{||\mathbf{w}(t)||}{||\mathbf{w}(0)||}\right).
\label{def le}
\end{equation}
In a $n$-dimensional problem we have $n$ Lyapunov exponents, each one referring to the divergence degree of specific directions of the system.
All of them form a set called Lyapunov Spectrum (LS), which usually are organized as:
\begin{equation}
\lambda_1 \geq \lambda_2 \geq \dots \geq \lambda_N.  
\label{LS}
\end{equation}
If the LLE $\lambda_1$ is positive then neighbor trajectories tend to diverge exponentially which implies a chaotic regime.
Due to the Liouville theorem, the Lyapunov spectrum of a Hamiltonian system is, as those considered below,
satisfies the relations (Pesin's theorem):
\begin{equation}
\lambda_i = -\lambda_{2N-i+1},\hspace{5mm}\lambda_{N+1}=\lambda_N=0.
\label{lypasymrel}
\end{equation}
Another important result relates the Kolmogorov-Sinai entropy and the Lyapunov spectrum. The former
measures exponential rate of information production in a dynamical system~\cite{ott} and
according to Pesin's theorem can be obtained as the sum of all positive Lyapunov exponents~\cite{eckmann}.

\subsection{Numeric determination of the LLE}

In the tangent map method, one considers the linearized form of Eq.~(\ref{flow}) around the point $x=x^*$:
\begin{equation}
\frac{d\mathbf{w}}{dt} = \mathbf{J}|_{x=x^*}\mathbf{w},
\label{mot_eq}
\end{equation}
where $\mathbf{J}$ is the Jacobian matrix of the vector function ${\mathbf F}({\bf x})$.
One then solves the original nonlinear system in Eq.~(\ref{flow}) and the linearized equations~(\ref{mot_eq}).
The steps for determining the Lyapunov exponent are~\cite{strelcyn,parker}:
\begin{enumerate}
\item For the nonlinear system~(\ref{flow}) impose an initial condition $\mathbf{x}_0$, and
an initial condition ${\mathbf w}_0=\pmb{\delta}_{0}$ for the linearized equations, with
$||\pmb{\delta}||=\epsilon$ and $\epsilon\ll1$.

\item Both differential equations are integrated for a time interval $T$.
This results in $\mathbf{x}_0 \rightarrow \mathbf{x}(T)$ and
$\pmb{\delta}_0\rightarrow\pmb{\delta}_1\equiv {\mathbf w}(T) $;
  
\item After each integration interval $T$, normalize the corresponding difference vector $\pmb{\delta}_k$ to $\epsilon$
and use the resulting vector as a new initial condition for solving the linearized equations;
 
\item The LLE is obtained from the average:
\begin{equation}
\lambda_1 = \displaystyle \frac{1}{KT} \sum_{k=1}^K \ln\frac{||\pmb{\delta}_k||}{\epsilon},
\label{avgle}
\end{equation}
where $K$ is chose in order to achieve convergence in the value of $\lambda_1$.
\end{enumerate}

By considering a solution $(\theta_i^*(t),p_i^*(t))$ of the equations of motion of the HMF model, the linearized equations
are obtained by plugging $\theta_i(t)=\theta_i^*(t)+\delta\theta_i(t)$ and $p_i(t)=p_i^*(t)+\delta p_i(t)$,
with small $\delta\theta_i(t)$ and $\delta p_i(t)$, into Eq.~(\ref{HMFeqmot}):
\begin{eqnarray}
\dot{\delta\theta_i} & = & \delta p_i,
\nonumber\\
\dot{\delta p_i} & = & -\left[ M_x^*\cos\theta_i^*+M_y^*\sin\theta_i^*\right]\delta\theta_i
\nonumber\\
 & & -\delta M_x\sin\theta_i^*+\delta M_y\cos\theta_i^*,
\label{eqsmothmflin2}
\end{eqnarray}
where the components of the magnetization are computed at the angles $\theta_i^*$ and are denoted $M_x^*$ and $M_y^*$
and
\begin{eqnarray}
\delta M_x(t)\equiv-\frac{1}{N}\sum_{j=1}^N\delta\theta_j(t)\sin\theta_j^*(t),
\nonumber\\
\delta M_y(t)\equiv\frac{1}{N}\sum_{j=1}^N\delta\theta_j(t)\cos\theta_j^*(t).
\label{defDeltas}
\end{eqnarray}
Both sets of equation in Eq.~(\ref{HMFeqmot}) and Eq.~(\ref{eqsmothmflin2}) must be solved simultaneously.

\begin{figure}[htbp]
\begin{center}
\includegraphics[scale=0.24]{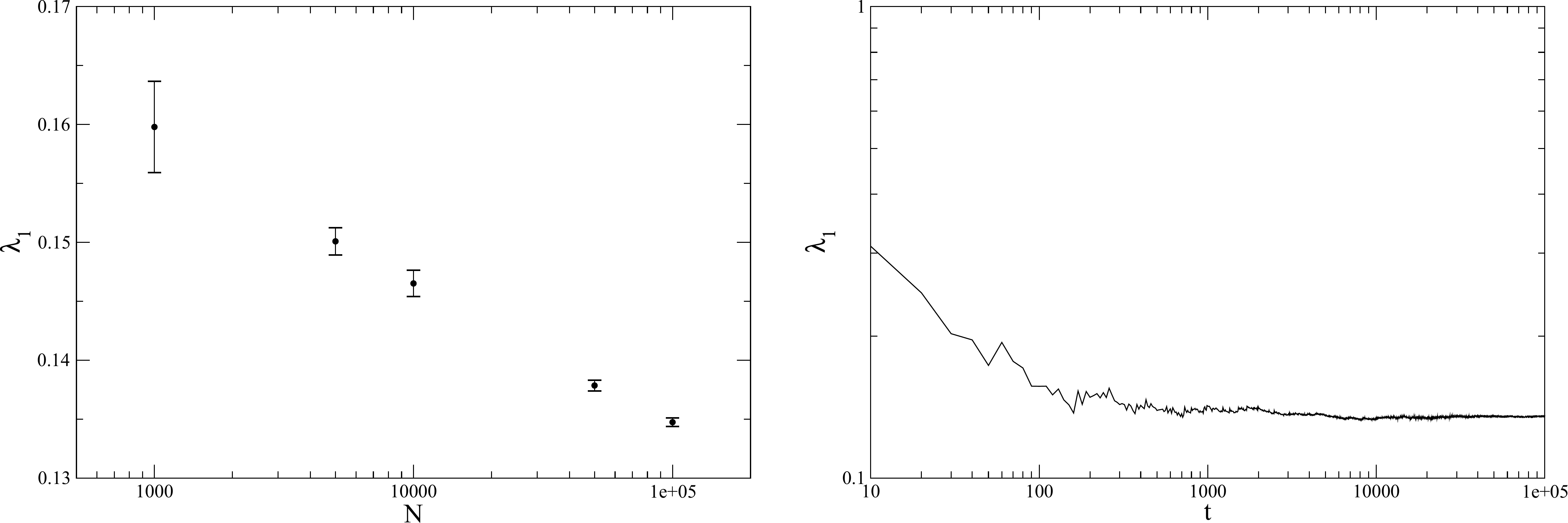}
\end{center}
\caption{Left Panel: the largest Lyapunov exponent $\lambda_1$ for a few values of $N$ ranging from $1000$ to $100\:000$
for the equilibrium state for an energy per particle of $e=0.5$ which corresponds to a magnetized state,
with the respective error bars obtained from $10$ different realizations of the initial conditions.
The parameters used are $T=10.0$ for the renormalization interval, numeric integration time step $\Delta t=0.05$ and total
integration time $t_f=10^5$. Right Panel: the value of $\lambda_1$ for $N=100\,000$
as a function of the total integration of time, showing good convergence.}
\label{lyaperrbarrs}
\end{figure}

In order to compute the LLE for very large values of $N$ the tangent map method was implemented in a parallel
code in graphic processing units~\cite{eu2} using a fourth-order symplectic integrator for both system~\cite{yoshida}.
Figure~\ref{lyaperrbarrs} shows the results for the computation of the LLE for some different values of $N$ and energy
per particle $e=0.5$. The error bars decrease rapidly with $N$, as expected. The right-panel of the same figure shows the
that convergence is achieved for a total simulation time   $t_f=10^5$. In all the results below we thus chose to use
the same parameter values and twice larger a value for $t_f$ to ensure proper convergence in all cases.

\subsection{An analytic estimate for the LLE}

The LLE for the HMF model was investigated numerically by Yamaguchi~\cite{yamaguchi}, and then latter
estimated by Latora, Rapisarda and Ruffo~\cite{latora}
from a random matrix approach of Parisi and Vulpiani~\cite{parisi}, and by Firpo~\cite{firpo2} using the differential geometry approach by
Pettini and collaborators~\cite{casetti,caiani,pettini2}. In the latter approach, the dynamics of the $N$ particle system is reformulated
in the framework of Riemannian geometry, where the trajectories correspond to geodesics of an underlying metric. Chaos then comes
from the instability of the geodesic flow, that at its turn depends on the properties of the curvature of the Riemannian manifold.
Chaos can also result from a parametric instability of the fluctuation of the curvature along
the system trajectory as represented in the manifold.

In order for the present paper to be self contained, we succinctly present here the the main results of
the geometrical approach to the computation of Lyapunov exponents and chaos from
Refs.~\cite{pettini2} and~\cite{casetti} (where the reader can find more details).
We consider a system of $N$ identical particles with unit mass and Hamiltonian:
\begin{equation}
H=\sum_{i=1}^N\frac{p_i^2}{2}+V({\bf r}_1,\ldots,{\bf r}_N),
\label{genericham}
\end{equation}
where ${\bf r}_i$ is the position vector of particle $i$, ${\bf p}_i$ its canonically conjugate momentum and $V$ the potential energy.
Considering two solutions $\overline{\bf r}_i(t)$ and ${\bf r}_i(t)$, $i=1,\ldots,N$, initially close to one another:
$\overline{\bf r}_i(t)={\bf r}_i(t)+\boldsymbol\xi_i(t)$.
The linearized equations of motion for the small variations $\xi_1(t)$ are given by:
\begin{equation}
\frac{d^2\boldsymbol\xi_i(t)}{dt^2}+\sum_{j=1}^N\frac{\partial^2 V}{\partial{\bf r}_i\partial{\bf r}_j}\cdot\boldsymbol\xi_j(t)=0.
\label{eqdiffgen}
\end{equation}
The maximal Lyapunov exponent is then obtained from:
\begin{equation}
\lambda=\lim_{t\rightarrow\infty}\frac{1}{2t}\ln\left(\frac{||\boldsymbol\xi(t)||^2}
{||\boldsymbol\xi(0)||}\right),
\label{lambdagen}
\end{equation}
where
\begin{equation}
||\boldsymbol\xi(t)||\equiv\sqrt{\boldsymbol\xi_1(t)^2+\cdots+\boldsymbol\xi_N(t)}.
\label{normxigen}
\end{equation}
The norm $\psi=||\boldsymbol\xi(t)||$ satisfies the equation:
\begin{equation}
\frac{d^2\psi(t)}{dt^2}+k(t)\psi(t),
\label{psieq}
\end{equation}
where $k(t)$ is a stochastic process describing the time evolution of the curvature along a trajectory in phase space.
Since the solutions of Eq.~(\ref{psieq}) are given in term of the averages $\langle\cdots\rangle$ over many realizations
of the stochastic process, Eq.~(\ref{lambdagen}) assumes the form:
\begin{equation}
\lambda=\lim_{t\rightarrow\infty}\frac{1}{t}
\label{lambdagen2}\frac{\langle\psi(t)\rangle}{\langle\psi(0)\rangle}.
\end{equation}
The solution for the average is obtained from a perturbative expansion in the amplitude of the fluctuations of the stochastic process,
which are small for large $N$. By introducing a smallness multiplicative parameter in the stochastic process as $k(t)\rightarrow\alpha k(t)$,
where $\alpha\ll1$, an supposing that fluctuations are delta correlated, a perturbative solution of Eq.~(\ref{psieq}) in powers of $\alpha$
can be obtained~\cite{pettini2,kampen}. In the case that $k(t)$ is a Gaussian process, this solution becomes exact.

An estimate of the LLE $\lambda_1$, with the extra assumption that the curvature along a trajectory is well represented by a Gaussian process,
was obtained in Ref.~\cite{casetti}, with very good agreement with numerical results for the Fermi-Pasta-Ulam model and
the 1D $XY$ model~\cite{pettini2}. The LLE is given in then given by~\cite{casetti}:
\begin{equation}
\lambda_1=\frac{\Lambda}{2}-\frac{2\kappa_0}{3\Lambda},
\label{LLEgeom}
\end{equation}
where
\begin{equation}
\Lambda=\left(2\sigma_k^2\tau+\sqrt{\frac{64}{27}\kappa_0^3+4\sigma_k^4\tau^2}\right)^{1/3},
\label{Lambdadef}
\end{equation}
\begin{equation}
\tau=\frac{\pi\sqrt{\kappa_0}}{2\sqrt{\kappa_0}\sqrt{\kappa_0+\sigma_k}+\pi\sigma_k},
\label{taudef}
\end{equation}
with $\kappa_0$ and $\sigma_k^2$ the average curvature and the fluctuations around its mean value, respectively, $\tau$ being a characteristic
time for the stochastic process. For the HMF model, Firpo obtained a closed form expression for the quantities $\kappa_0$ and $\sigma_k$~\cite{firpo2},
such that the (Ricci) scalar curvature in the Riemannian manifold is given by $\kappa_R=M^2$. The next step consists to take
$\kappa_0=\langle M^2\rangle_\mu$, i.~e.\ the microcanonical average of $M^2$. The variance of the curvature fluctuations in the microcanonical ensemble
was obtained in~\cite{firpo2} as:
\begin{equation}
\sigma_k^2=\langle\delta^2 k_R\rangle_c\left(1+\frac{\beta^2}{2}\langle\delta^2 k_R\rangle_c\right)^{-1},
\label{sigmahmf}
\end{equation}
where $\langle\delta^2 k_R\rangle_c$ is the variance of the fluctuations in the canonical ensemble:
\begin{equation}
\langle\delta^2 k_R\rangle_c=4M\frac{\partial M}{\partial\beta},
\label{canfluchmf}
\end{equation}
with $M$ given by the solution of Eq.~(\ref{eqdisthmf}). It is worth noting that even if $k(t)$ is not Gaussian, the expressions
above remain valid up to first order in $\alpha$.

Previous results showed that in the non-magnetized phase the Lyapunov exponent tends to zero
as $N^{-1/3}$ obtained in numerical simulations in Ref.~\cite{latora} and predicted theoretically in~\cite{firpo2}.
In the ferromagnetic phase, a more complicate picture emerges. Manos and Ruffo observed numerically a transition from a weak
to a strong chaoticity regimes at low energy~\cite{manos}, and related it to the time dependence of the phase of the magnetization vector,
which becomes strongly time dependent around the same energy (a more detailed explanation of this point is given in Ref.~\cite{rochamarcos}).
A critical exponent for the LLE was predicted by Firpo~\cite{firpo2}
in the vicinity of the second order phase transition for $e<e_c$ in the form
\begin{equation}
\lambda_1\propto(e_c-e)^\xi,
\label{criticalexp}
\end{equation}
with an exponent $\xi=1/6$. Ginelli and collaborators obtained a different value $\xi=1/2$ from numerical results,
the same critical behavior as the magnetization~\cite{ginelli}. Below we obtain a value of $\xi$ close to the theoretical value
by considering much higher values of $N$.

\section{Results}
\label{sec3}

\begin{figure}[htbp]
\begin{center}
\includegraphics[scale=0.24]{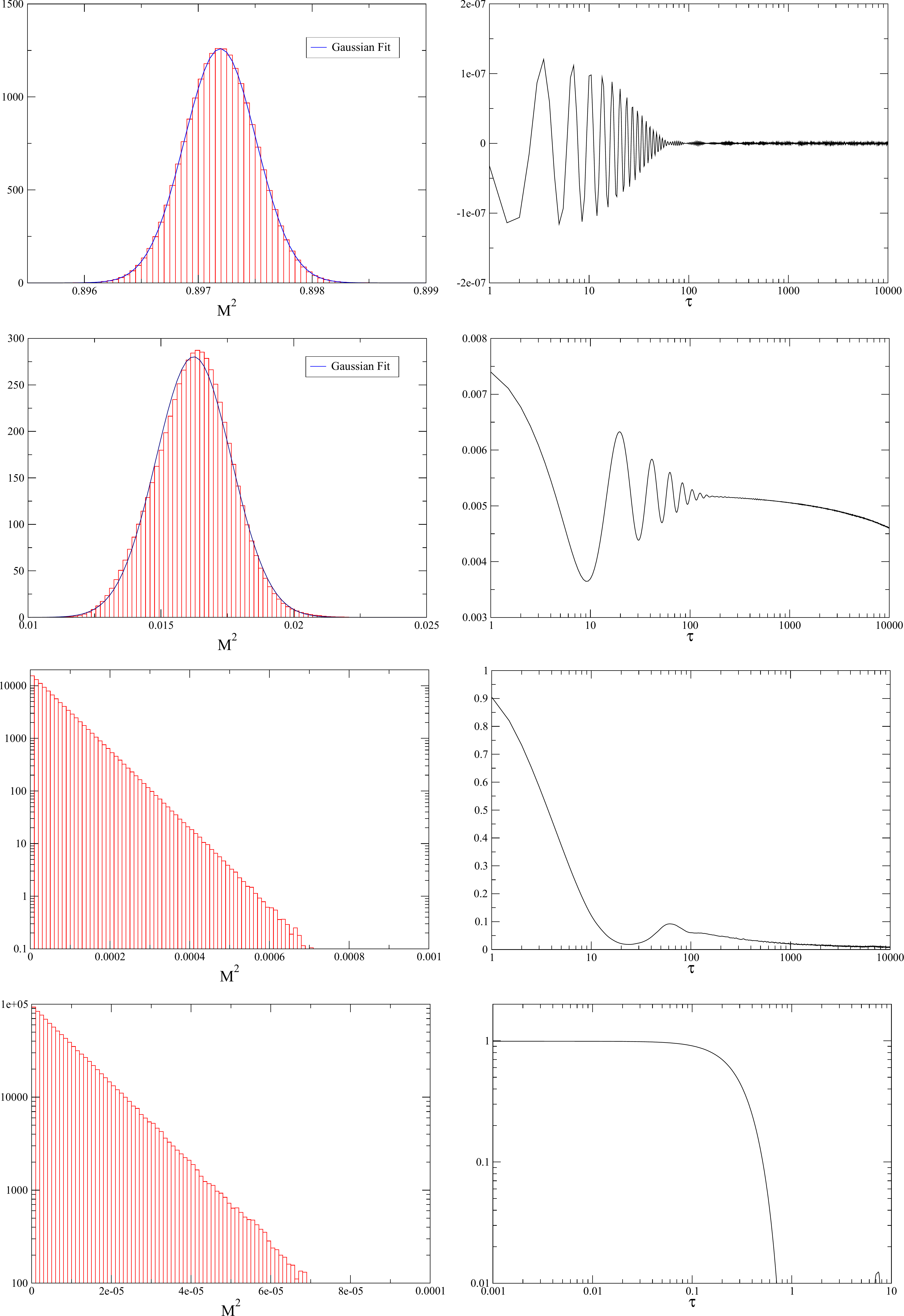}
\end{center}
\caption{(Color online) a) Distribution (histogram) of fluctuations of the curvature $\kappa_R=M^2$ for the equilibrium state with energy per particle $e=0.1$,
$N=100\,000$. The continuous line is a least squares fit with a Gaussian distribution.
b) Correlation $\langle M(t_0-\tau)M(t_0)\rangle-\langle M\rangle^2$ as a function of $\tau$.
c) and d) Same as (a) and (b) for e=0.74. e) Log plot of the distribution of fluctuations of $\kappa_R$ for $e=0.8$ where
an exponential distribution is clearly visible. f) Correlations for $e=0.8$.
g) same as (a) with $e=5.0$. g) Same as (b) for $e=5.0$ In all cases the total simulation time is $t_f=10^5$,
integration time step $\Delta t=0.05$, except (g) and (h) where $t_f=10^4$ and $\Delta t=10^{-2}$.}
\label{distcorr}
\end{figure}

\begin{figure}[htbp]
\begin{center}
\includegraphics[scale=0.25]{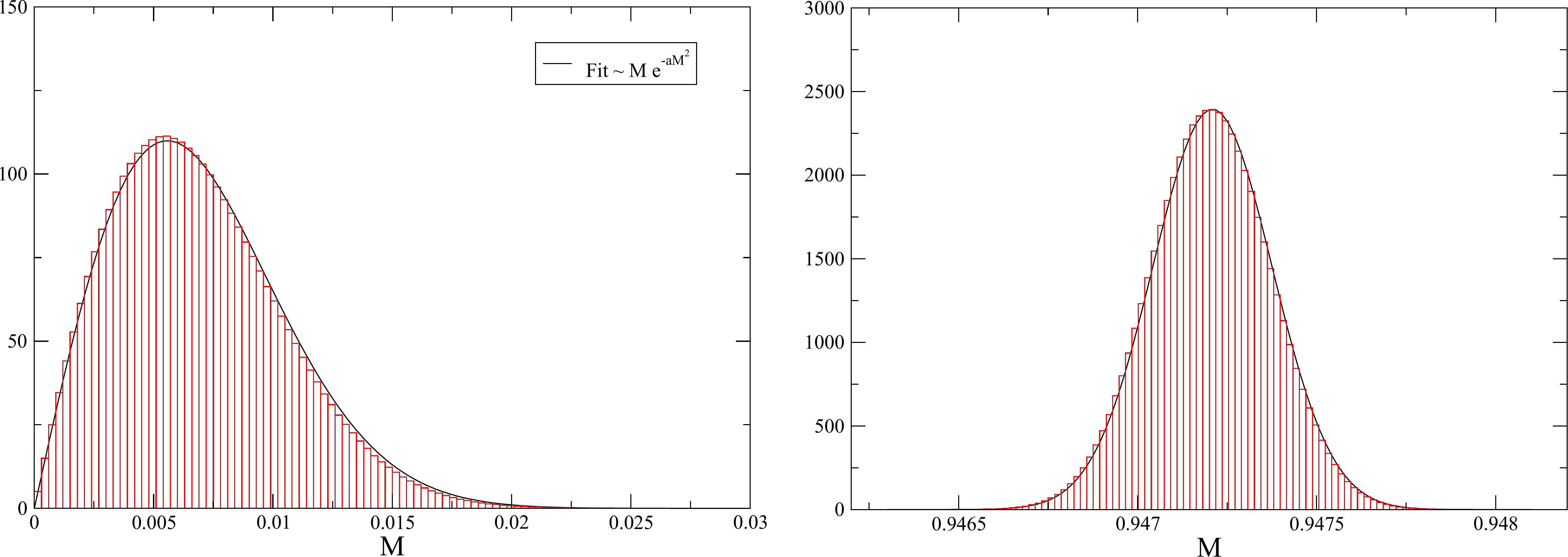}
\end{center}
\caption{(Color online) Left Panel: Distribution of values of the magnetization $M$ for the equilibrium state with $e=0.8$. The continuous line
is a least squares fit of the expression $bM\exp\left(-aM^2\right)$, with $a$ and $b$ constants.
Right Panel: Same as in the left panel but for $e=0.1$. The continuous line is a fitting of a Gaussian function
$a\exp\left(-a(M-\langle M\rangle)^2\right)$.}
\label{mdist}
\end{figure}

The first point to consider is whether the fluctuations of the curvature, i.~e.\ of $M^2$ for the HMF model, can be modeled by
an uncorrelated Gaussian process, as considered in Refs~\cite{firpo2,casetti}. Figure~\ref{distcorr} shows the distributions of the
fluctuations of the curvature $\kappa_R=M^2$ for a few values of energy, and the correlation function for the fluctuations
$\langle M(t_0-\tau)^2M(t_0)^2\rangle-\langle M^2\rangle^2$ for a few energy values. In the ferromagnetic state the fluctuations
are well described by a Gaussian distribution, but the correlation time, i.~e.\ the time for correlations to be negligible, can be very large.
The correlation time is small only at higher energies.
In the homogeneous phase, correlations of the fluctuations of the curvature are also non-negligible,
and their distribution is non-Gaussian quite close to an exponential function.
In fact in this case it is more natural to expect that the fluctuations of the magnetization components are Gaussian rather than those of $M^2$,
thus explaining the form of the distributions in Fig.~\ref{distcorr}e and~\ref{distcorr}g.
The distributions for the values of $M$
are given in Fig.~\ref{mdist}. Below the critical energy the distribution is Gaussian, while above the phase transition it is well described
by a function of the form $bM\exp(-aM^2)$, with $a$ and $b$ constants.
In obtaining Eqs.~(\ref{LLEgeom}--\ref{taudef}) the central assumption was that fluctuations are delta correlated. This is clearly valid only
for higher energies, where as shown below the predicted $N^{-1/3}$ scaling of the LLE is observed.
Deviations from the theoretical predictions are thus expected for lower energies due to strong correlations in the fluctuations of the curvature.

\begin{figure}[htbp]
\begin{center}
\includegraphics[scale=0.3]{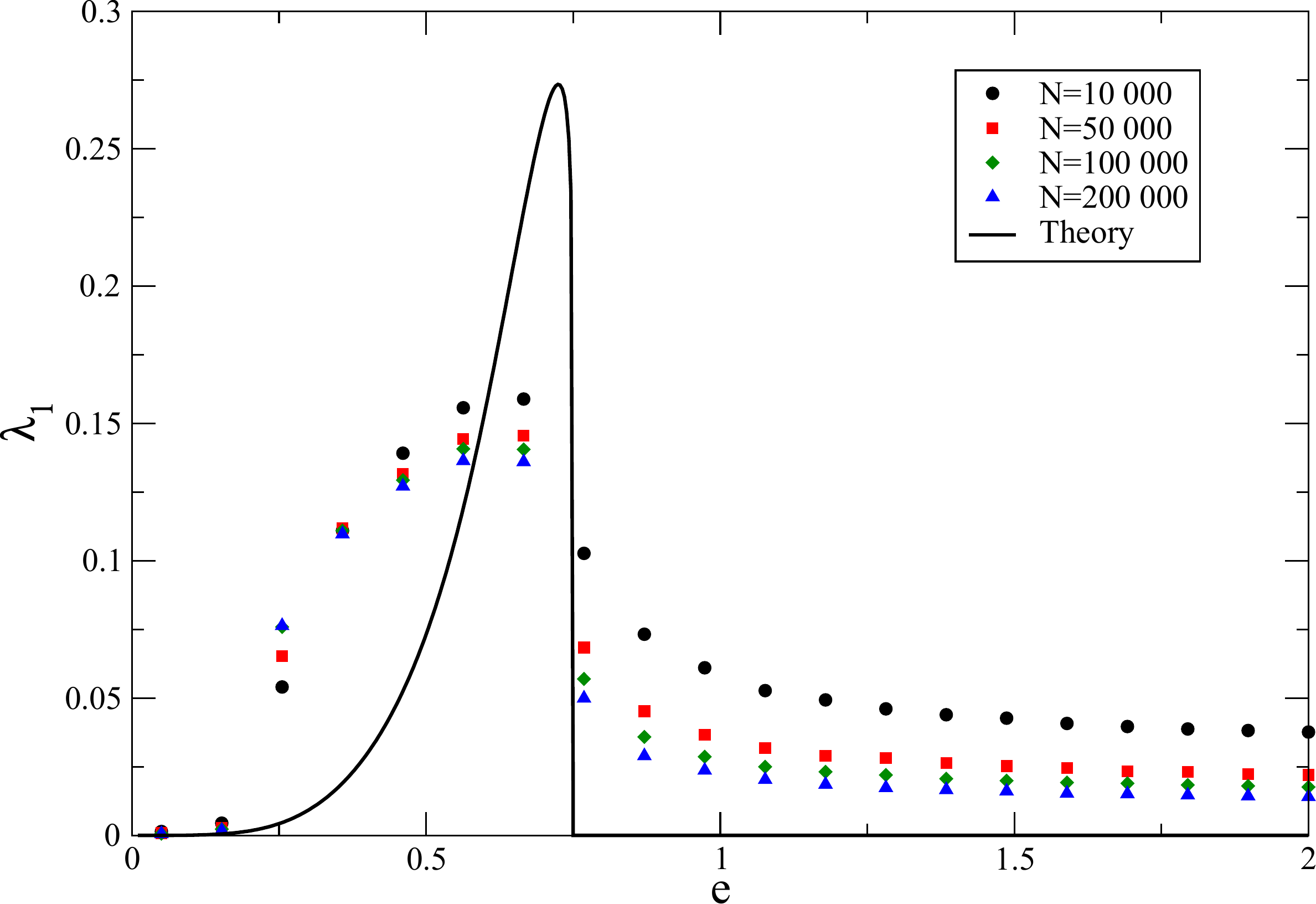}
\end{center}
\caption{(Color online) Largest Lyapunov exponent $\lambda_1$ as a function of energy per particle
$e$ for a few values of $N$. The continuous line is the theoretical result from Ref.~\cite{firpo1},
and the vertical dotted line indicates the position of the second order phase transition.
The parameters used in the simulation are $\Delta t=0.05$ for the numeric integration time step, $T=10.0$
as the time interval between two renormalizations and total simulation time $t_f=2\times10^5$.}
\label{LE_transition}
\end{figure}

\begin{figure}[htbp]
\begin{center}
\includegraphics[scale=0.23]{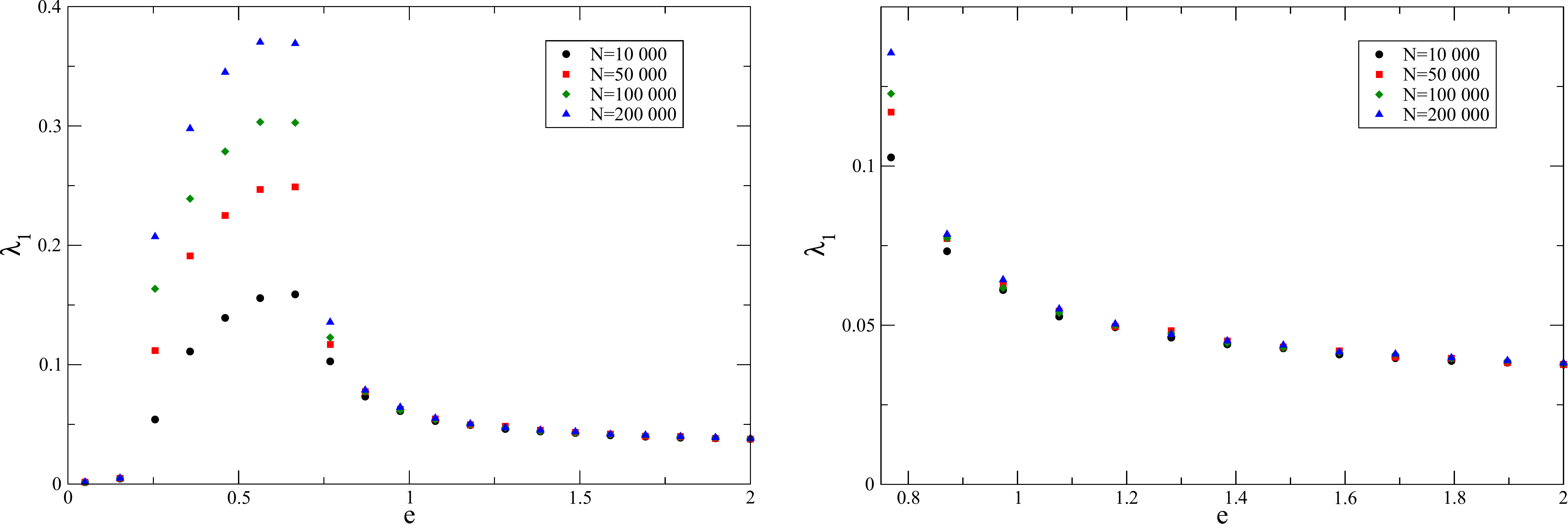}
\end{center}
\caption{(Color online) Left panel: same as in the left panel of Fig.~\ref{LE_transition} but with $\lambda_1$ rescaled by $(N/10\,000)^{-1/3}$.
Right panel: zoom over the energy range $(0.75,2.0)$.}
\label{LE_transition_scal}
\end{figure}

\begin{figure}[htbp]
\begin{center}
\includegraphics[scale=0.3]{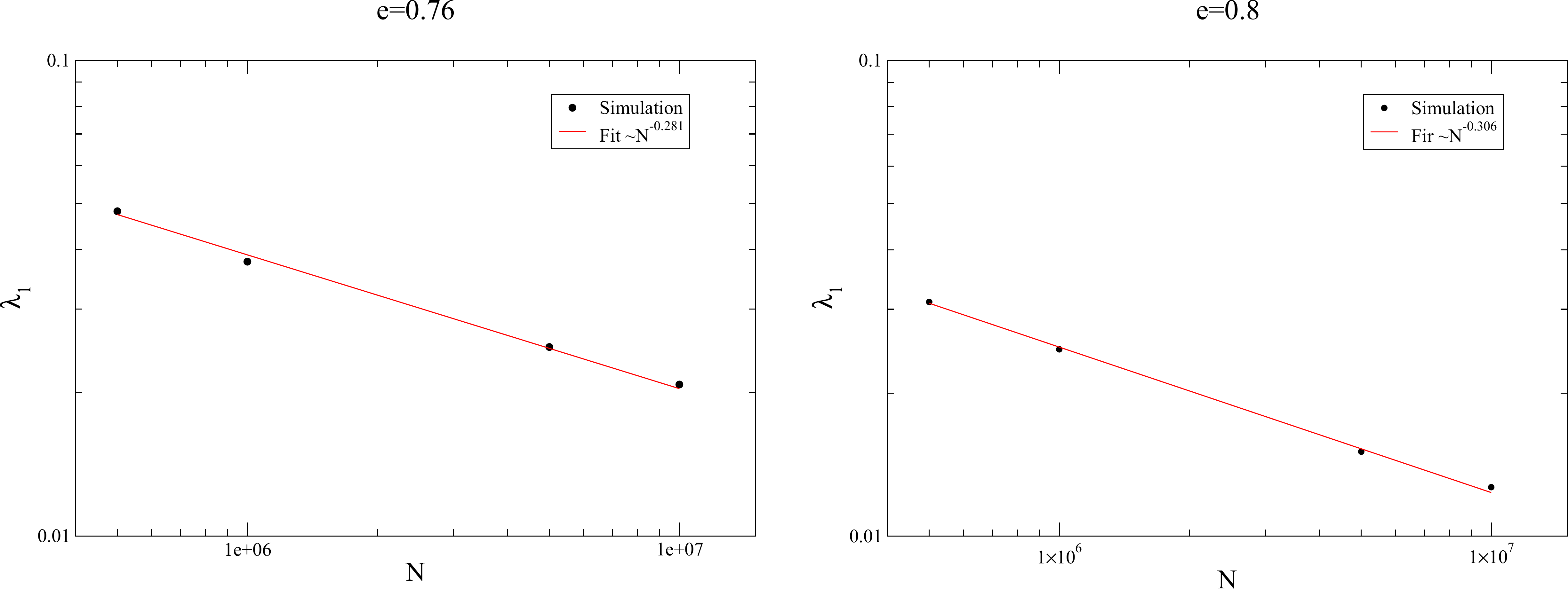}
\end{center}
\caption{Largest Lyapunov exponent $\lambda_1$ for $e=0.76$ and $e=0.8$, with values of $N$ ranging from $N=500\,000$ up to $N=10\,000\,000$ 
for one single realization. The continuous line is a least-squares fit of a power law $N^\gamma$. We note that for larger values of $N$ the
exponent $\gamma$ approaches the theoretical value of $1/3$.}
\label{scaling2}
\end{figure}

\begin{figure}[htbp]
\begin{center}
\includegraphics[width=70mm]{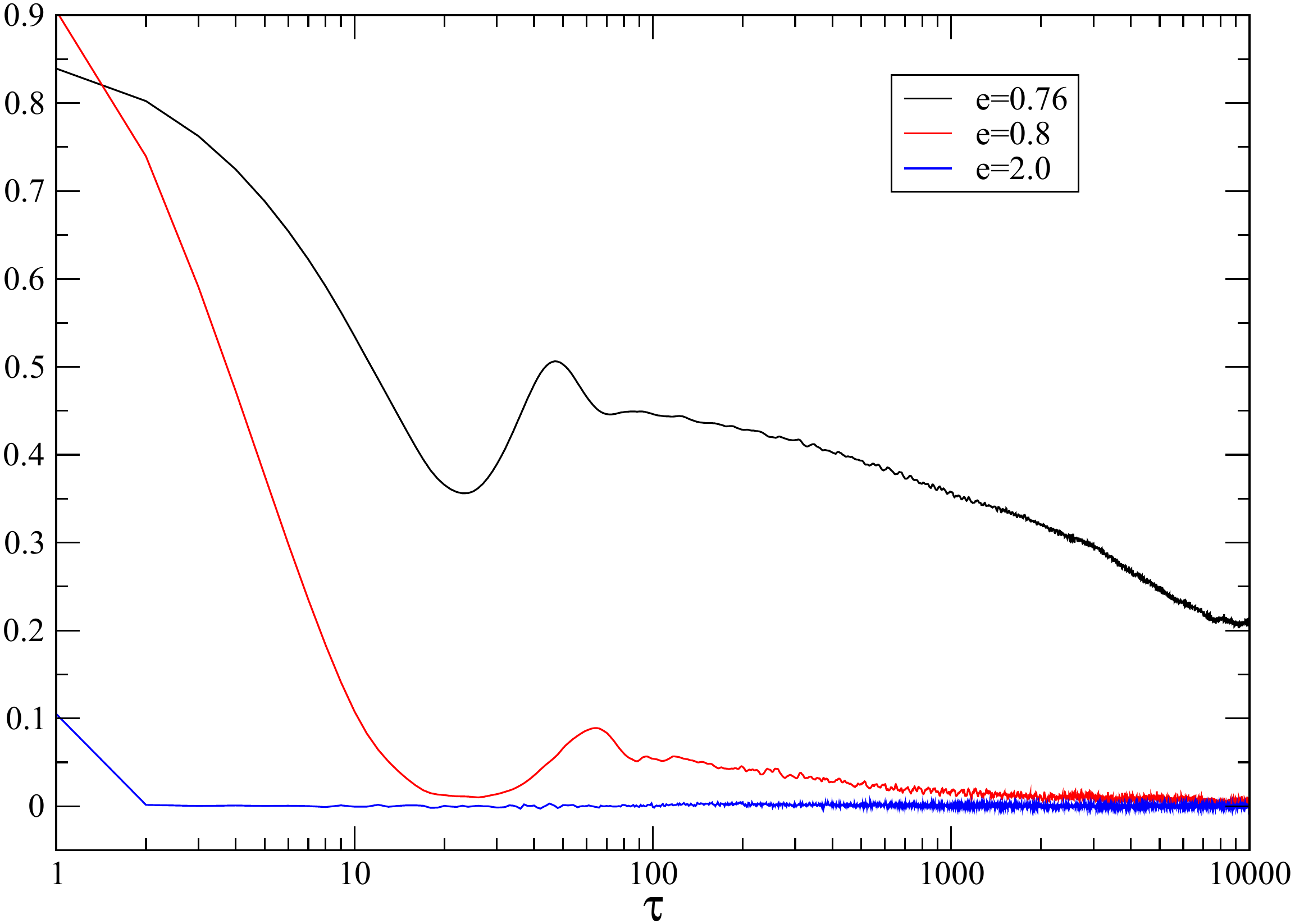}
\end{center}
\caption{Correlation function $\langle M(t_0+\tau)^2M(t_0)^2\rangle$ for the fluctuations in the curvature $\kappa_R=M^2$
for a few values of $e$ and $N=100\:000$.}
\label{corrXe}
\end{figure}

Figure~\ref{LE_transition} shows the plot of the LLE $\lambda_1$ as a function of energy for some values of $N$, alongside
the theoretical prediction of Ref.~\cite{firpo2}. The parameters used in the numeric integration are
$T=10.0$ and $\delta t=0.05$ which are used in all simulations below unless explicitly stated.
In the ferromagnetic phase the theoretical results agree only qualitatively
with numerical results, predicting a maximum of the LLE for an energy below the critical energy $e_c$, but not its position,
and also that $\lambda_1$ goes to zero at the phase transition.
The left panel in Fig.~\ref{LE_transition_scal} shows a reasonable
data collapse if the exponent are rescaled by $N^{-1/3}$, that nevertheless becomes not so good for energies closer to the phase
transition as seen on the left panel of Fig.~\ref{LE_transition_scal}.
\begin{figure}[htbp]
\centering
\includegraphics[scale=0.23]{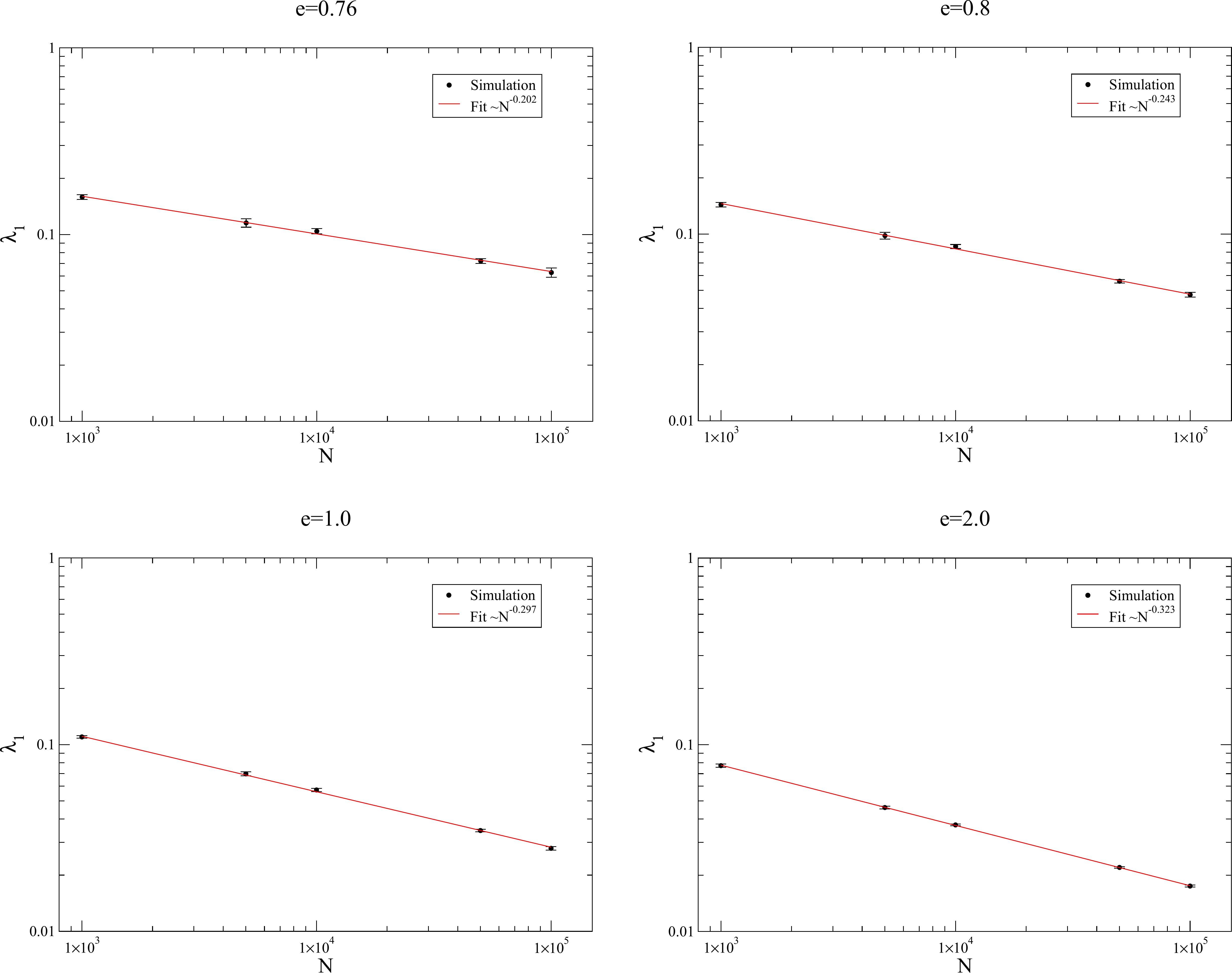}
\caption{(Color online) Largest Lyapunov exponent $\lambda_1$ for a few energy values $e=0.76$, $0.8$, $1.0$ and $2.0$. The
error bars were obtained from 10 realizations for each value of $N$. The continuous line is a chi-square fit of a power law in $N$.
The numeric parameters are the same as in Fig.~\ref{LE_transition}.}
\label{LE_scale}
\end{figure}
Figure~\ref{LE_scale} shows $\lambda_1$ as a function of $N$ for some energy energy values in the non-magnetized state.
The predicted $N^{-1/3}$ scale is observed far from the phase transition. Nevertheless for higher values of $N$ we slowly
approach the $N^{1/3}$ scaling as shown in Fig.~\ref{scaling2} for the energy $e=0.8$.
This can be explained by the fact
that the fluctuations of the Riemannian curvature are not delta correlated close to the phase transition as seen from the
correlation functions in Fig.~\ref{corrXe}.
It is also important to note that for non-Gaussian fluctuations the solution of Eq.~(\ref{psieq}) is only valid
at order $\alpha^2$, and therefore is more accurate for smaller $\alpha$ and equivalently greater $N$.

\begin{figure}[htbp]
\centering
\includegraphics[scale=0.23]{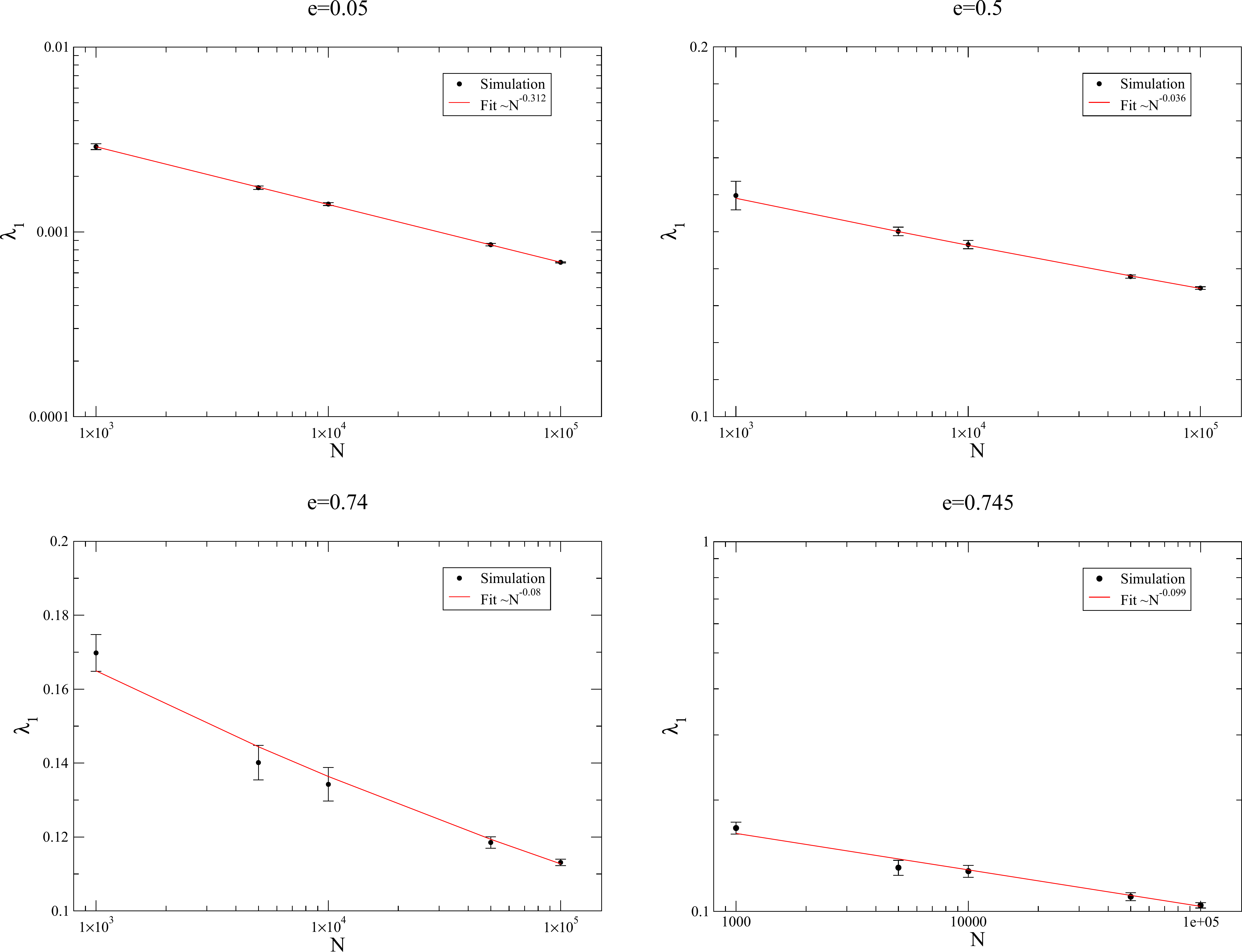}
\caption{Largest Lyapunov exponent $\lambda_1$ for a few energy values $e=0.05$, $0.5$, $0.74$ and $0.745$.
The numeric parameters are the same as in Fig.~\ref{LE_transition}.}
\label{LE_scale2}
\end{figure}

For the ferromagnetic phase, Figure~\ref{LE_scale2} shows the LLE $\lambda_1$ as a function of $N$ for a few energy values.
The scaling of the LLE with $N$ is close to $N^{-1/3}$ for very low energies while it is much
slower for energies above $e_w\approx0.15$ and below the critical energy.
Manos and Ruffo studying the same system observed a
transition from weak to strong chaos at the same energy $e_w$, such that below it the LLE is much smaller
and scales as $N^{-1/3}$, while no results for the scaling of the LLE with $N$ were obtained for $e>e_w$~\cite{manos}.
The same authors using the generalized alignment indices method~\cite{skokos} showed that at this energy the fraction of chaotic orbits
of the system increases rapidly from a very low (less than 1\%) to a very large value (close to 100\%).
As a consequence, the convergence of the LLE to zero in the mean-field limit is non-uniform, which characterizes two distinct energy intervals.
For $e<e_w$ (weak chaos) the LLE rapidly tends to zero, while having a significant positive value for $e_w<e<e_c$ (strong chaos)
up to relatively high values of N.
\begin{figure}[htbp]
\begin{center}
% OK!!!!
\includegraphics[scale=0.3]{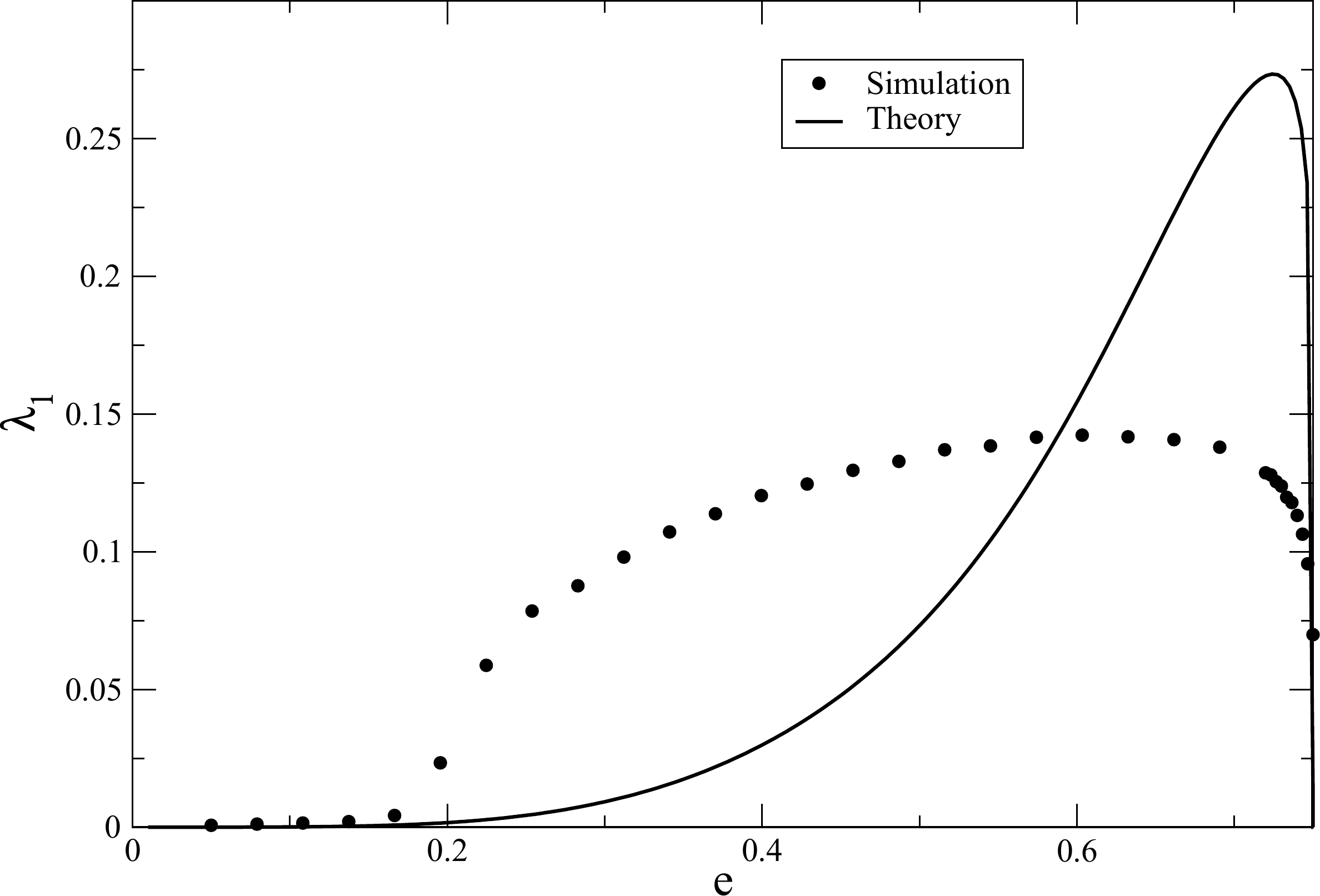}
\end{center}
\caption{Largest Lyapunov exponent $\lambda_1$ as a function of energy for $N=100\,000$ for $e<e_c=0.75$ from numerical simulations.
The (smooth) transition from weak to strong chaoticity is clearly visible at $e\approx0.15$. The continuous line is the theoretical prediction.}
\label{lyapbelow}
\end{figure}

\begin{figure}[htbp]
\begin{center}
\includegraphics[scale=0.3]{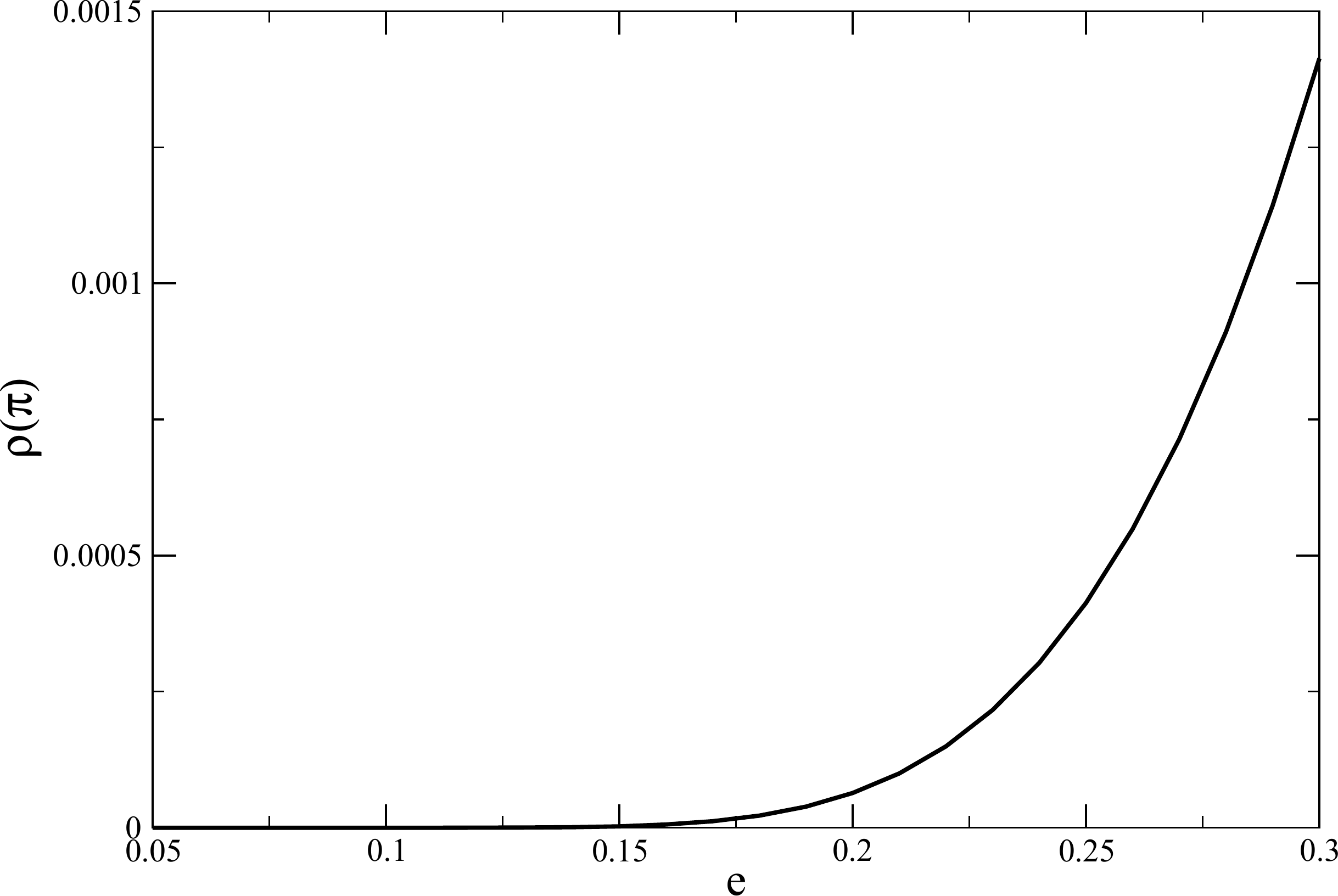}
\end{center}
\caption{Value of the spatial distribution function $\rho_{eq}(\theta)$ in Eq.~(\ref{spatialdist}) at $\theta=\pi$ as a function of energy.
We observe that $\rho(\pi)$ becomes non-negligible very close to the (smooth) transition from weak to strong chaos}
\label{border}
\end{figure}

The transition from weak to strong chaos can be explained from the equilibrium properties of the system.
The equilibrium spatial distribution obtained by integrating $f_{eq}$ in Eq.~(\ref{eqdisthmf}) over the momentum, is given by
\begin{equation}
\rho_{eq}(\theta)=C_N\exp\left(\beta M\cos(\theta)\right),
\label{spatialdist}
\end{equation}
with $C_N$ a normalization constant. In Eq.~(\ref{spatialdist}) the maximum of $\rho_{eq}(\theta)$ occurs at $\theta=0$ by a choice of the origin
for the angles. The values of the spatial distribution at $\theta=\pi$ as a function of energy are shown in Fig.~\ref{border}.
As already pointed out by Manos and Ruffo~\cite{manos}, the transition from weak to strong chaos occurs at the energy value $e_w$
when $\rho_{eq}(\pi)$ attains a significant value
and particles start to cross at the border $\theta=\pi$, causing a time variation of the phase of the magnetization due to
asymmetries in the fluctuations of the distribution in Eq.~(\ref{spatialdist}) which is valid for $N\rightarrow\infty$.
Indeed, the equations of motion in Eq.~(\ref{HMFeqmot}) for any particle in the system can be written as the equation of a pendulum:
\begin{equation}
\ddot\theta=-M\sin(\theta+\phi),
\label{pendeqmot}
\end{equation}
where $M=\sqrt{M_x^2+M_y^2}$ and $\phi=\arctan{M_y/M_x}$. If the phase $\phi$ is time independent the solutions of Eq.~(\ref{pendeqmot})
are non-chaotic, while having a positive Lyapunov exponent for a time varying phase, which occurs significantly
in the strong chaos energy interval.
This point is explored in more detail in Ref.~\cite{rochamarcos}.

\begin{figure}
\begin{center}
\includegraphics[width=140mm]{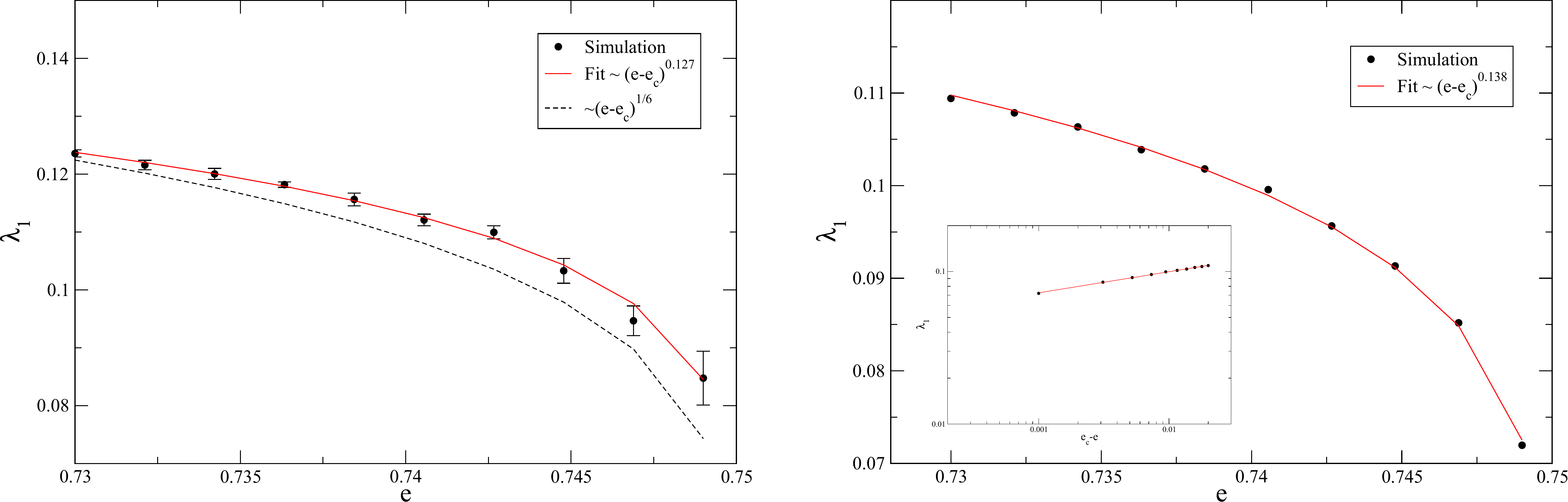}
\end{center}
\caption{Left Panel (Color Online): Largest Lyapunov exponent $\lambda_1$ as a function of energy close to the phase transition for $N=100\,000$.
The continuous line is a least squares fit of a power law $(e-e_x)^b$ with $b=0.127$ and
the dashed line is proportional to the theoretical function with $b=1/6$ drawn for comparison.
The error bars were obtained from 10 different realizations for each energy.
The simulation parameters are $T=10.0$, $t_f=2\times10^5$ and $\Delta t=0.05$.
Right Panel: Same as the left panel but with $N=1000\,000$. The inset shows a log-log plot of $\lambda_1$ as a function of $e_c-e$.
The exponent in the fit is $b=0.138$ slighter close to the theoretical value.}
\label{critbelow2}
\end{figure}

As a last result, we investigate the possible critical behavior of the LLE for energies close to $e_c$ from below as the theoretical prediction
in Eq.~(\ref{criticalexp}), by numerically determining $\lambda_1$. Although theoretically predicted no numerical verification has been
obtained previous to the present work.
The results are shown in Fig.~\ref{critbelow2} for $N=100\:000$ and $N=1000\:000$ and some energy values.
The fitting of Eq.~(\ref{criticalexp}) is very good for both values of $N$,
with an exponent close to the theoretical value $1/6$. Small deviations are possibly due to important correlations
in the fluctuations of $\kappa_R=M^2$, which become more important close to the phase transition, as discussed above.
The difference with respect to the exponent $\xi\approx1/2$ obtained in Ref.~\cite{ginelli} can be explained by our longer
simulation times and higher values of $N$ which were made feasible by a massively parallel implementation of our numeric code.

\section{Concluding Remarks}
\label{sec4}

This paper addressed the study of chaoticity in the HMF model from the determination of LLE. This paradigmatic model has been widely
used in the literature to understand the behavior and some properties of long range interacting systems.
Our numerical implementation CUDA allowed to investigate the LLE for a wide range of energies, and values of $N$
as large as $2\times10^7$. The size of the system has been essential to describe
the main characteristics features of the exponents. For the homogeneous phase ($e \geq 0.75$,) at all energies, it was shown clearly
that the exponents scales with the system size as $N^\beta$ with $\beta$ approaching $-1/3$, the theoretical predicted value. Close
to the phase transition we must go to higher values of $N$ in order to observe the expected scaling.
This comes from non negligible self-correlations in time of the fluctuations of the scalar curvature used in the geometric approach
for the theoretical determination of the LLE. For energies below the transition energy we observe two different scaling for the
LLE: for energies below $e_w\approx0.15$ the LLE scales approximately with $1/N^{1/3}$, while for $e_w<e<e_c$ the exponent of the scaling
is much smaller than $1/3$. This is explained first by non-negligible correlations in the fluctuations of the curvature of the underlying
Riemannian manifold and second by the coupling of the motion of individual particles to a time varying phase of the magnetization.

We also confirmed numerically the existence of a critical exponent associated to the Lyapunov exponent as defined in Eq.~(\ref{criticalexp}).
The value we have obtained for this exponent is $\xi\approx0.138$ with is reasonably close to the predicted theoretical value of $1/6$, and far from
the value of $1/2$ obtained in Ref.~\cite{ginelli}.
With respect to the former, this difference is explainable by the fact that the stochastic process representing the Riemannian curvature on the underlying manifold
in not delta correlated, as shown in Fig.~\ref{distcorr}d. Our parallel implementation of the algorithm for computing the LLE allowed a significant
improvement in the accuracy of the numerical results, which possibly explains the variance with the result in~\cite{ginelli}.

Whether such a critical exponent
also occurs for other long range interacting systems is an open question that requires to be investigated. A similar but much computationally demanding
study for the self-gravitating ring model~\cite{sota} is the subject of ongoing work.

Finally we close this section by pointing out that the theoretical results of Firpo~\cite{firpo2},
although based on some necessary simplifying assumptions with respect to the geometrical approach of Pettini and collaborators
yields results quite often close to our numerical findings. The discrepancies are then explained when those assumptions are not
valid, as for instance when the fluctuations of the curvature are non-negligible.

\section{Acknowledgments}

MAA and TMRF would like to thank M.-C.~Firpo for fruitful discussions. TMRF was partially financed by CNPq (Brazil) and
LAMF was financed by CAPES (Brazil).

\end{document}